\documentstyle[12pt]{article}
\def\be{\begin{equation}}
\def\ee{\end{equation}}
\def\ba{\begin{array}{c}}
\def\ea{\end{array}}

\def\ben{$$}
\def\een{$$}
\begin{document}

\titlepage

 \begin{center}
{\Large \bf New exact solutions for polynomial oscillators in large
dimensions}

\end{center}

\vspace{5mm}

 \begin{center}

Miloslav Znojil

\vspace{3mm}

{\small \it Nuclear Physics Institute AS CR, 250 68 \v{R}e\v{z},
Czech Republic\\

e-mail: znojil@ujf.cas.cz}

\vspace{5mm}

Denis Yanovich and Vladimir P. Gerdt

 \vspace{3mm}

{\small \it Lab. Inf. Tech., Joint Institute for Nuclear Research
(JINR) 141980, Dubna, Moscow Region, Russia\\
 email: yan@jinr.ru  and gerdt@jinr.ru}



\end{center}

\subsection*{Abstract}

A new type of exact solvability is reported. Schr\"{o}dinger equation
is considered in a very large spatial dimension $D \gg 1$ and its
central polynomial potential is allowed to depend on ``many" (= $2q$)
coupling constants. In a search for its bound states possessing an
exact and elementary wave functions $\psi$ (proportional to a
harmonic-oscillator-like polynomial of a freely varying, i.e., not
just small, degree $N$), the ``solvability conditions" are known to
form a complicated nonlinear set which requires a purely numerical
treatment at a generic choice of $D$, $q$ and $N$. Assuming that $D$
is large we discovered and demonstrate that this problem may be {\em
completely} factorized and acquires an amazingly simple exact
solution at all $N$ and up to $q = 5$ at least.

\vspace{3mm}

PACS {03.65.Ge}

\newpage

\section{Introduction}

The key motivation of our present study of polynomial oscillators
lies in the well known fact that the {\em majority} of quantitative
predictions in nuclear, atomic, molecular and condensed matter
physics must rely on a more or less purely numerical model. The
completely non-numerically tractable quantum systems are rare though,
at the same time, useful and transparent (cf., e.g., the description
of vibrations in molecules mimicked by harmonic oscillator).
Polynomial oscillators may be taken, in this setting, as lying
somewhere on a borderline between the two regimes.

The first indications of a breakdown of the traditional separation
between the numerical and analytic models in quantum mechanics came
with the emergence of certain ``incompletely analytic" (now we call
them quasi-exactly solvable, QES) polynomial oscillator models.
Unfortunately, their separate discoveries at the beginning of the
last quarter of the twentieth century \cite{Hautot} --
\cite{singular} have all been understood and accepted as a mere
formal curiosity. As an interesting confirmation of analogies with
classical mechanics for the charged and shifted oscillator in two and
three dimensions \cite{Hautot}, as a peculiar singularity in a
universal continued-fraction algorithm for sextic oscillators
\cite{Singh}, as an exceptional case in the standard infinite-series
solution of the corresponding differential Schr\"{o}dinger equations
\cite{Flessas} -- \cite{singular} etc.

The defining property of the QES models (being solvable just for a
part of the complete set of their coupling constants and/or energies)
re-acquired a new meaning only after their Lie-algebraic
re-interpretation \cite{Turbiner} which revealed their genuine
mathematical closeness to the exactly solvable (ES) Hamiltonians
\cite{Miller}. Subsequently,  in physics, a multitude of their
relationship to the apparently different models has been revealed in
a way summarized, e.g., in the detailed Ushveridze's monograph
\cite{Ushveridze}.

In spite of the unique success of the {\em mathematical} QES
formulae, a number of difficulties  remained connected with their
{\em practical} applications and applicability. One of the key
reasons (and differences from the current ES models) is that the {\em
explicit construction} of the multiplet QES energies {\em remains
purely numerical}. Indeed, these values must be computed as roots of
an $N-$dimensional secular determinant so that the difference between
the variational, ``generic $N=\infty$" rule in Hilbert space seems
only marginally simplified by the QES construction of any QES
multiplet with $N \gg 1$.

The main purpose of our present study is related precisely to the
latter point. Our idea may be explained, briefly, as an application
of perturbative philosophy assuming that the {\em spatial} dimension
$D$ is large. In this spirit we are going to consider a general
Magyari's \cite{Magyari} QES Hamiltonian $H^{(q,N)}(D)$ (at a fixed
dimension $D$ of the space of coordinates and with a chosen size $N$
of its secular determinants, see below for a more detailed
explanation). Finally we {\em construct a set} of its specific
approximations $H_0^{(q,N)}(\infty)$ with errors proportional to
$1/D$.

The {\em exact solvability} of the latter Hamiltonians $H_0^{(q,N)}$
emerged as an utterly unexpected result of our (originally, fully
numerical) calculations. Our presentation of it starts from a concise
review of the concepts of exact solvability in section~2. Section~3
will then mention a few specific formal merits of transition to the
domain of the large dimensions $D$ which simplifies the general
Magyari's secular determinants considerably. The core of our message
appears in section 4 where in the domain of $D \gg 1$, the separate
families of the polynomial oscillators (numbered by the integer $q=
1, 2 $ and $3$) are studied in detail and shown to lead to the closed
solutions. Section 5 outlines the possibilities of an extension of
these results to $q=4$ and $5$ while section 6 summarizes and
discusses some possible practical consequences.

\section{Exactly solvable oscillators: A  brief review}

\subsection{Harmonic oscillator and the like:
 All bound states are elementary}

One of the most exceptional exactly solvable models in quantum
mechanics is the central harmonic oscillator in $D$ dimensions.
Its so called superintegrability (the term coined by Pavel
Winternitz \cite{Winternitz}) makes its partial differential
Schr\"{o}dinger equation
 \be
\left (
 -
\frac{\hbar^2}{2m} \triangle + \frac{1}{2}\,m\,\Omega^2\,|\vec{x}|^2
\right ) \Psi(\vec{x}) = \varepsilon\, \Psi(\vec{x}) \
 \label{HOgeneric}
 \ee
solvable by the separation of variables in several systems of
coordinates. The most common cartesian choice may be recommended
for the first few lowest spatial dimensions $D$ only
\cite{Fluegge3}. In contrast, the separation in spherical system
remains equally transparent at any $D$ because it reduces eq.
(\ref{HOgeneric}) to {\em the same} ordinary (so called radial)
differential equation
 \be
 \left[-\,\frac{d^2}{dr^2} + \frac{\ell(\ell+1)}{r^2} +
\omega^2r^2
 \right]\, \psi(r) =
E\, \psi(r)
 \label{rad}\
 \ee
with  $r = |\vec{x}| \in (0,\infty)$,  $E
={2m\varepsilon}/{\hbar^2}$ and $ \omega={m\Omega}/{\hbar}>0$. In
this language we have $\ell = \ell_L= L + {(D-3)}/{2}$ where $L =
0, 1, \ldots $ \cite{BG}. At each $L$ the energy levels are
numbered by the second integer,
 \be
E = E_{n,L}=\omega\,(2n+\ell_L+3/2), \ \ \ \ \ \
 \ \ \ \ \ \
 n,L = 0,
1, \ldots\  \label{closeden}.
 \ee
The wave functions with quadratic $ \lambda(r)=\omega\,r^2/2>0$ and
minimal $N = n+1$ in
 \be
 \psi_{n,L}(r) =r^{\ell+1}\,
 e^{-\lambda(r)}\,\sum_{m=0}^{N-1}h_m\,r^{2m} \
  \label{ansatz}
 \ee
are proportional to an $n-$th Laguerre polynomial \cite{FluLag}. In
Hilbert space, their set is complete: This may be explained via
oscillation theorems and characterizes the harmonic oscillator as
exceptional. In such a setting, one should recollect all the similar
(i.e., Coulomb and Morse) exactly solvable potentials, but one need
not mention them separately as long as they are formally equivalent
to our harmonic oscillator example after a simple change of variables
\cite{CKS}.

\subsection{Sextic oscillators:
Many bound states are elementary}

For many phenomenological purposes, even the shifted harmonic
oscillator forces $V^{(HO)}(r) =V^{(HO)}(0)+\omega^2 r^2$ are not
flexible enough. Fortunately, there exists an immediate QES
generalization of the harmonic oscillators revealed by Singh et al
\cite{Singh}. Let us summarize briefly this type of solvability as a
construction which starts from a replacement of the free constants in
$V^{(HO)}(r)$ by their polynomial descendants of the first order in
$r^2$,
 \be
 \omega \longrightarrow W(r)
  = \alpha_0 +\alpha_1r^2 \,, \ \ \ \ \ \ \
 V^{(HO)}(0) \longrightarrow  G_{-1}+ G_0r^2 = U(r)\,.
 \label{omega}
 \ee
After the conventional choice of $G_{-1}=0$ this trick gives the
general sextic potential
 \begin{equation}
 V^{(sextic)}(r) =U(r)+ r^2W^2(r)=
 g_0\,r^2+g_1\,r^4 + g_{2}\,r^{6}\,
 \label{SExta}
 \end{equation}
all three couplings of which are simple functions of our initial
three parameters and {\it vice versa},
 \be
 \ \ \ g_{2}= \alpha_1^2 > 0,
 \ \ \ g_{1}= 2\,\alpha_0\alpha_1,
 \ \ \  g_{0}= 2\,\alpha_0^2+G_0\,.
 \label{mapa}
 \ee
The resulting Schr\"{o}dinger bound state problem cannot be solved
in closed form. Nevertheless, we may postulate the polynomiality
of the wave functions $ \psi^{(sextic)}_{n,L}(r)$ {\em for a
finite multiplet} (i.e., $N-$plet) of the wave functions. For this
purpose it is necessary to lower the number of the freely variable
couplings by the specific constraint
 \be
 G_0=-\alpha_0^2-\alpha_1(4N+2\ell+1), \ \ \ \ \ \ N \geq 1
 \label{lowers}\ .
  \ee
For each$N-$plet, the polynomial solutions (\ref{ansatz}) are made
exact by the $N-$dependent  QES condition (\ref{lowers}). The choice
of a WKB-like (i.e., quartic) exponent
 \be
  \lambda(r) =
 \frac{1}{2}{\alpha}_0 r^2 + \frac{1}{4}{\alpha}_1 r^4\,
 \label{qrawS}
 \ee
guarantees their physical normalizability. The ansatz (\ref{ansatz})
transforms then the differential Schr\"{o}dinger equation into a
linear algebraic definition of the unknown $N-$plet of coefficients
$h_m$. An incomplete solution is always obtained for a mere finite
set of the levels $n \in (n_0, n_1, \ldots,n_{N-1})$. In contrast to
the harmonic oscillator, the QES solvability is based on the $L-$ and
$N-$ dependent constraint (\ref{lowers}) so that, generically, the
elementary QES multiplet exists in a single partial wave only.

\subsection{General
harmonic-oscillator-like bound states}

The explicit energy formula (\ref{closeden}) for harmonic oscillator
was replaced by an implicit definition in the preceding paragraph
which gives the sextic QES energies {\em in the purely numerical
form}, viz., as zeros of the Singh's secular determinant of a certain
tridiagonal $N$ by $N$ matrix \cite{Ushveridze}. In this sense, a
very natural further extension of the Singh's QES construction exists
and has been described by Magyari \cite{Magyari}. Its description may
start from the more consequent replacement (\ref{omega}), constructed
from even polynomials of degree $2q$,
 \ben
 V^{(q)}(r) =U^{(q)}(r)+ r^2[W^{(q)}(r)]^2\,,
 \ \ \ \ \ \ \  U^{(q)}(r) = G_0r^2 +  G_1r^4 +\ldots + G_{q-1}r^{2q}
 \,,
  \een
  \be
 \ \ \ \ \ \ \
W^{(q)}(r)
  = \alpha_0 +\alpha_1r^2 + \ldots + \alpha_q\,r^{2q}\,
 \label{greeSEx}
 \ee
This formula re-parametrizes the polynomial
 \begin{equation}
 V^{[q]}(r) =
 g_0\,r^2+g_1\,r^4 + \ldots + g_{2{q}}\,r^{4{q}+2}\,
 \ \ \ \ \ \ \ g_{2q}= \gamma^2 > 0
 \label{geSExt}
 \end{equation}
and specifies the one-to-one correspondence between the two sets
of couplings,
 \ben
\{g_0, \ldots, g_{2q} \} \Longleftrightarrow \{G_0, \ldots,
G_{q-1}, \alpha_0, \ldots, \alpha_{q} \}\,
 \een
where $g_{2q}= {{\alpha}_{q}}^2$, $g_{2q-1} =g_{2q-1}
(\alpha_{q},\alpha_{q-1})= 2\,{\alpha}_{q-1}\,{\alpha}_{q}, \ldots$
or, in opposite direction, ${\alpha}_{q}=\sqrt{g_{2q}}\equiv \gamma
>0$, ${\alpha}_{q-1}=g_{2q-1}/{(2\alpha_q)}$ etc.

At a generic $q=1, 2, \ldots$, equation (\ref{qrawS}) must be further
modified in such a way that $r\,W(r)\equiv Z'(r)$,
 \be
 \lambda^{(q)}(r) =
 \frac{1}{2}{\alpha}_0 r^2 + \frac{1}{4}{\alpha}_1 r^4 +
\ldots + \frac{1}{2q+2} {\alpha}_{q} r^{2q+2}
 \label{qraw} \label{Jost}\,.
 \ee
With $\alpha_q > 0$, one verifies that
 \ben
 \psi^{(physical)}(r) \approx e^{-\lambda^{(q)}(r)+{\cal O}(1)},\
 \ \ \ \   r \gg 1\,
 \een
which means that the correct bound-state ansatz
 \begin{equation}
\psi(r) = \sum_{n=0}^{N-1}\, h_n^{(N)} \,r^{2n+\ell+1}\,{\rm
exp}\left [ -\lambda^{(q)}(r)\right ]
 \label{ana}
 \end{equation}
converts our radial equation (\ref{rad}) + (\ref{geSExt}) into an
equivalent linear algebraic problem
\begin{equation}
\hat{Q}^{[N]}\,\vec{h}^{(N)}=0\,.
 \label{tridSE}
 \end{equation}
Its closer inspection reveals that this problem is overcomplete,
i.e., its matrix is {\em non-square} and asymmetric,
 \be
  \hat{Q}^{[N]}=
 \left(
  \begin{array}{lllllll}
 B_{0} & C_0&  & & && \\
 A_1^{(1)}&B_{1} & C_1&    &&& \\
 \vdots&&\ddots&\ddots&&&\\
 A_q^{(q)}& \ldots& A_q^{(1)} &
 B_{q} & C_q&    & \\
 &\ddots&&&\ddots&\ddots&\\
 &&A_{N-2}^{(q)}& \ldots & A_{N-2}^{(1)}&
 B_{N-2} & C_{N-2} \\
 &&&A_{N-1}^{(q)}& \ldots & A_{N-1}^{(1)}&
 B_{N-1} \\
 &&&&\ddots&\vdots&\vdots\\
 &&&&&A_{N+q-2}^{(q)}&A_{N+q-2}^{(q-1)}\\
 &
 &&&&&A_{N+q-1}^{(q)}\\
 \end{array}
 \right ).
 \label{hatted}
 \ee
Its elements depend on the parameters in bilinear manner,
\begin{equation}
\begin{array}{c}
   C_n = (2n+2)\,(2n+2\ell+3),\ \ \ \ \ \
  B_n = E-\alpha_0\,(4n+2\ell +{3})
\\
 A_n^{(1)} = -\alpha_1\,(4n+2\ell+1) + \alpha_0^2 -g_0, \ \ \ \ \ \ \
 A_n^{(2)} = -\alpha_2\,(4n+2\ell-1) + 2 \alpha_0\alpha_1 -g_1,  \\
 \ldots,\\
 A_n^{(q)} = -\alpha_q\,(4n+2\ell+3-2q) +
 \left (\alpha_0\alpha_{q-1}+ \alpha_1\alpha_{q-2}+ \ldots
  +\alpha_{q-1}\alpha_0
 \right ) -g_{q-1},
 \\
\ \ \ \ \ \ \ \ \ \ \ \ \ \ \ \ \
  \ \ \ \ n = 0, 1, \ldots \ .
  \end{array}
  \label{elem2}
   \end{equation}
At {\em any} fixed and finite dimension $N =1, 2, \ldots$ the
non-square system (\ref{tridSE}) is an over-determined set of $N+q$
linear equations for the $N$ non-vanishing components of the vector
$\vec{h}^{(N)}$.

\subsection{Changes of variables and an extension of applicability
of the harmonic-oscillator-like constructions}

At $q=0$, equations (\ref{hatted}) degenerate back to the recurrences
and define the well known harmonic oscillator states. As already
mentioned, their additional merit lies in the coincidence of their
polynomial part with the current Laguere polynomials.

At $q=1$ we return to the sextic model where the ''redundant" last
row fixes one of the couplings and where we are left with a
diagonalization of an $N$ by $N$ matrix which defines the $N-$plet of
the real QES energies in principle. In such a setting, an important
piece of an additional encouragement results from the well known
possibility of a definition of new QES Hamiltonians by a change of
variables $r \to x$ and $\psi(r) \to x^{const} \phi(x)$ in the
Schr\"{o}dinger equation \cite{CKS,Leach}. Even when considered just
in the most elementary power-law form, this change is defined by the
prescription \cite{classif}
 \ben
 r^{2j}\ \ \ \longrightarrow \ \ \
 x^{\delta(k)},
 \ \ \ \ \
 \delta(k) = 2\frac{j+1}{k}-2, \ \ \ \
 k = 1, 2, \ldots, 2q+2,
 \een
and extends the class of the ''interesting" potentials not only in
the well known $q=0$ case (see the open possibility of a transition
to the completely solvable Coulombic potential as mentioned above)
but also in the $q=1$ model where the following four equivalent
potentials may be distinguished and numbered by the above index $k$
attached to them as their second superscript,
 \be
\ba
 V^{(q=1,k=1)}(r)= { a
r^{2} + b r^{4} +  r^{6} }
\\
 V^{(q=1,k=2)}(r)=   {
a r^{-1} + b r +  r^{2} }
\\
 V^{(q=1,k=3)}(r)=  a
r^{-4/3} + b r^{-2/3} +  r^{2/3}
\\
 V^{(q=1,k=4)}(r)=  {  a
r^{-3/2} + b r^{-1} + c r^{-1/2} }\ .
\end{array}
\label{nesix}
 \ee
The situation is more complicated at $ q > 1$. The counting of
parameters and equations indicates that unless one broadens the class
of potentials, only a very small multiplet of bound states may remain
available in closed form \cite{Leach}. Still, the same elementary
change of variables enables us to extend the set of the partially
solvable potentials to the six-member family at $q=2$,
 \be
\ba
 V^{(q=2,k=1)}(r)= { a
r^{2} + b r^{4} + c r^{6} + d r^{8} + r^{10} }
\\
 V^{(q=2,k=2)}(r)=   {
a r^{-1} + b r+ c r^{2} + d r^{3} +  r^{4} }
\\
 V^{(q=2,k=3)}(r)=  a
r^{-4/3} + b r^{-2/3}+ c r^{2/3} + d r^{4/3} +  r^{2}
\\
 V^{(q=2,k=4)}(r)=  {  a
r^{-3/2} + b r^{-1}+ c r^{-1/2} + d r^{1/2} + r }
\\
 V^{(q=2,k=5)}(r)=  {  a
r^{-8/5} + b r^{-6/5}+ c r^{-4/5} + d r^{-2/5} + r^{2/5} }
\\
 V^{(q=2,k=6)}(r)=  {  a
r^{-5/3} + b r^{-4/3}+ c r^{-1} + d r^{-2/3} + f r^{-1/3} }\
\end{array}
\label{nedekix}
 \ee
etc (cf., e.g., paper \cite{classif} where the similar lists of the
partially solvable potentials have been displayed up to $q=4$). In
this way, the availability of the exact solutions for all these
various forces might offer a new inspiration, say, in some
phenomenological considerations and models and/or for their
perturbative analyses and some more detailed large$-D$ calculations.

\section{Solvability of polynomial oscillators
at the large spatial dimensions}

Up to now, our attention has been concentrated upon the structure of
the QES wave functions. From the point of view of the evaluation of
the energies, the main dividing line between the solvable and
unsolvable spectra is in fact marked by the distinction between the
closed $q=0$ formulae and their implicit QES form at $q=1$. The
transition to the next $q=2$ may be perceived as merely technical. At
all $q \geq 1$, the difficulties grow with $N$. In such a setting we
noticed the emergence of certain simplifications at $D \gg 1$.

\subsection{Difficulties arising at $q \geq 1$}

At any $D$, the last row in eq. (\ref{tridSE}) decouples from the
rest of the system. At any $q>1$ it may treated as a constraint
which generalizes eq. (\ref{lowers}),
 \be
 g_{q-1}  = -\alpha_q\,(4n+2\ell+3-2q) +
 \left (\alpha_0\alpha_{q-1}+ \alpha_1\alpha_{q-2}+ \ldots
  +\alpha_{q-1}\alpha_0
 \right ).
 \label{redef}
 \ee
The insertion of this explicit definition of the coupling
$g_{q-1}$ simplifies the lowest diagonal in $\hat{Q}^{[N]}$,
 \be
 A_n^{(q)} = 4\,\gamma\,(N+q-n-1).
 \ee
Since $ A_{N+q-1}^{(q)}=0$ we may omit the last line from eq.
(\ref{hatted}) and drop the ''hat" $\hat{\,}$ of the diminished
matrix $ \hat{Q}^{[N]}$. This gives eq. (\ref{tridSE}) in the more
compact form
 \begin{equation}
Q^{[N]}\,\vec{h}^{(N)}=0\
 \label{ridiSE}
 \end{equation}
where the size of the non-square matrix $Q^{[N]}$ is merely $(N+q-1)$
by $N$. Unfortunately, the new equation is still purely numerical,
with an exception of the simplest special case with $q=0$ where no
coupling is fixed and where the energies themselves are given by the
explicit formula (\ref{redef}). At $q=0$ also the recurrences for
coefficients of the wave functions may be solved in a compact form.

As already mentioned, the $q=1$ version of eq. (\ref{ridiSE})
degenerates to the secular equation
 \be
 \det Q^{[N]}=0.
 \label{secular}
 \ee
This is a purely numerical problem at all the larger $N \geq 5$.
Still, one coupling is fixed by eq. (\ref{redef}) and only the
$N-$plet of energies must be calculated as the real zeros of
secular polynomial.

At $q\geq 2$ the $q$ independent (and mutually coupled) $N$ by $N$
secular determinants must vanish at once \cite{Dubna}. With an
auxiliary abbreviation for the energy $E = -g_{-1}$ this means that
at least one of the couplings is always energy-dependent and its
value must be determined numerically. In the other words, our
non-square matrix $Q^{[N]}=Q^{[N]}(g_{-1}, g_0, \ldots,g_{q-2})$ will
annihilate the vector $\vec{h}^{(N)}$ if and only if all its $q$
arguments are determined in a deeply nonlinear and self-consistent,
mostly purely numerical manner.

\subsection{Simplifications arising at $D \gg 1$ \label{reasons}}

For a clear understanding of what happens at $D \gg 1$, let us pick
up the $q=0$ model $V^{[q]}(r)$ and re-consider its
coordinate-dependence in the $D \geq 1$ regime. We discover a quick
growth of the minimum of the effective potential, occuring at a
fairly large value of the coordinate $R =R(D)=
[\ell(\ell+1)/\omega^2]^{1/4}\gg 1$. Its Taylor expansion
 \be
 \frac{\ell(\ell+1)}{r^2} +
\omega^2r^2 =
 2\omega^2R^2 + 4\,\omega^2(r-R)^2-\frac{4}{R}(r-R)^3+ \ldots\,
 \label{effekt}
 \ee
reveals that the shape of the effective potential is $R-$ and
$D-$independent. Near the minimum, also the cubic and higher
corrections become negligible. This implies that the shifted harmonic
oscillations characterize now the local solutions very well. In
particular we have the wave functions
 \be
 \psi_0 \sim e^{-\omega\,(r-R)^2}, \ \ \ \ \ \ \
 \psi_1 \sim (r-R)\,e^{-\omega\,(r-R)^2}
 , \ \ \ldots, \ \ \ \ \ \ \ r \approx R\,
 \ee
and re-derive also the leading-order degeneracy of the spectrum and
its equidistance in the next order,
 \be
 E_0= 2\omega^2R^2 + 2\omega + \ldots, \ \ \ \ \ \ \
 E_1= 2\omega^2R^2 + 6\omega+\ldots, \ \ \ldots\,.
 \label{aprox}
 \ee
The agreement of this approximate formula with the available exact
spectrum (\ref{closeden}) is amazing.

Considerations which have led to this agreement may be applied as a
guide to the large-$D$ description of the low-lying states in {\em
any} phenomenological input potential well $V(r)$. In such a context,
a skeptical question is due. Once we have the exact formula, why
should we search for its alternative (re-)derivation? The reply will
follow from our forthcoming results. We shall see that a
qualitatively different new source of $D \gg 1$ simplification will
emerge in {\em all} the exactly solvable polynomial $q < \infty$
models.

\subsection{Magyari equations in the large$-D$ regime}

In the above $D \gg 1$ construction, little information can be
extracted from the wave functions themselves. Although we
Taylor-expanded the effective potential, we did not make any use of
the information about the wave functions. In this sense, we are now
going to demonstrate the feasibility of the approach where the {\em
guaranteed} polynomiality of the wave functions will play a key role.

In our original differential eq. (\ref{rad}) as well as in all its
$q>0$ generalizations, the numerical value of the spatial dimension
$D$ will be assumed large. This will simplify our matrix
re-arrangement (\ref{ridiSE}) of this problem with the matrix
elements (\ref{elem2}) re-written as the {\em linear} functions of
$D$,
\begin{equation}
\begin{array}{c}
   C_n = (2n+2)\,(2n+2L+D),\ \ \ \ \ \
  B_n = -g_{-1}-\alpha_0\,(4n+2L+D),
\\
 A_n^{(k)} = -g_{k-1} -\alpha_k\,(4n+2L+D-2k) +
 \left (\alpha_0\alpha_{k-1}+  \ldots
  +\alpha_{k-1}\alpha_0
 \right ),\\
\ \ \ \ \ \ \ \
 k = 1, 2, \ldots, q-1,\
 \ \
  \ \ \ \ \
  \ \ \ \ n = 0, 1, \ldots , N+q-2\ .
  \end{array}
  \label{elemih2}
   \end{equation}
Besides the $D-$independent $ A_n^{(q)} = A_n^{(q)[0]} $ which
remains unchanged, we shall preserve the dominant components of the
matrix elements,
  \ben
     C_n^{[0]} = (2n+2)\,D,\ \ \ \ \ \
  B_n^{[0]} = -g_{-1} -\alpha_0\,D,
\ \ \ \ \ \ \
 A_n^{(k)[0]} = -g_{k-1} -\alpha_k\,D \ ,k < q.
  \label{emih2}
   \een
Then we re-scale the coordinates and, hence, coefficients,
 \be
 h^{(N)}_n=p_n/\mu^n\ .
 \label{muna}
 \ee
By the choice of the parameter $\mu$ we are free to achieve that
the uppermost and the lowest diagonals are just a re-ordering of
each other,
 \be
 \frac{2D}{\mu} = \tau = 4\,\gamma\,\mu^q.
 \ee
This exhausts the freedom and fixes the $D-$dependence of our
scaling,
 \be
 \mu=\mu(D) = \left ( \frac{D}{2\gamma} \right )^{1/(q+1)},
 \ \ \ \ \ \
 \tau=\tau(D) = \left ( 2^{q+2}\,D^{q}\,\gamma
 \right )^{1/(q+1)}
 .\label{mu}
 \ee
The recipe replaces the energies and couplings $\{g_{-1}, g_0,
\ldots,g_{q-2}\}$ by the new parameters $\{ s_1, s_2,\ldots,
s_q\}$ in linear way,
 \be
 g_{k-2}=
 -\alpha_{k-1}D-\frac{\tau}{\mu^{k-1}}\,s_k,
 \ \ \ \ \ \ \
 k = 1, 2, \ldots, q\ .
 \label{recipe}
 \ee
In the leading-order approximation, our re-scaled Magyari
equations read
 \begin{equation}
 \label{trap}
 \left( \begin{array}{cccccc}
 s_1 & 1 & & & & \\
 s_2 & s_1 & 2 & &  & \\
 \vdots&  &\ddots & \ddots & & \\
 s_q & \vdots &  &s_1 & N-2& \\
 N-1& s_q & &  & s_{1} & N-1 \\
 &N-2& s_q & & \vdots& s_{1}  \\
 &&\ddots&\ddots&&\vdots\\
 &&& 2 & s_q&  s_{q-1}   \\ &&&& 1 & s_q
\end{array} \right)
 \left( \begin{array}{c}
 {p}_0\\
 {p}_1\\
\vdots \\
 {p}_{N-2}\\
 {p}_{N-1}
\end{array} \right)
=
0\ .
 \end{equation}
We shall now study their solutions.

\subsection{Inspiration: New closed formulae at  $q=0$ }

At $q=0$, as already mentioned, the energies are unique functions
of $N$ and do not exhibit any unexpected behaviour. Still, it is
instructive to extract the leading-order result for wave
functions. The $q=0$ version of eq. (\ref{trap})
 \begin{equation}
 \label{trapnol}
 \left( \begin{array}{cccccc}
 N-1 & 1 & & && \\
  & N-2 & 2 &  &&\\
 &  &\ddots & \ddots && \\
 &&& 2 & N-2&     \\&& && 1 & N-1
\end{array} \right)
 \left( \begin{array}{c}
 {p}_0\\
 {p}_1\\
\vdots \\
 {p}_{N-2}\\
 {p}_{N-1}
\end{array} \right)
=
0\
 \end{equation}
defines the (up to the normalization, unique) Taylor coefficients
 \be
 p_n=(-1)^n\,p_0\,
 \left (
 \ba
 N-1\\
 n
 \ea
 \right )\ .
 \ee
We may appreciate that they do not carry any round-off error and
that the related leading-order wave functions possess the
elementary form
 \be
 \psi_{n,L}(r) =r^{\ell+1}\,
 \left(
 1-\frac{r^2}{\mu}
 \right )^{n}\,
 e^{-\lambda\,r^2}\,, \ \ \ \ \ n = 0, 1, \ldots
  \label{anine}\,.
 \ee
At the same time one must be aware that only the leading-order part
of (\ref{anine}) is to be compared with the available exact $D <
\infty$ result. In particular, the presence of a degenerate nodal
zero is an artifact of the zero-order construction. All this
experience may serve as a guide to the less transparent $q>0$ cases.

\section{New partially solvable models with $q\geq 1$,
 any $N$ and large $D \gg 1$}

In the way inspired by eq. (\ref{nedekix}), one may move beyond $q=0$
and $q=1$ and transform the decadic forces into their quartic
equivalents etc. Paper \cite{classif} may be consulted for details
which indicate that the study of any potential $V(r)$ {\em which is a
polynomial in the powers of the coordinate $r$} may be replaced by
the study of its present Magyari's or ''canonical" QES representation
$V^{(q)}(r)$ at a suitable integer $q$. In addition, we shall also
restrict our attention to the domain of large $D$.

\subsection{Guide: Sextic QES oscillator with $q=1$ and any $N$}

Starting from the first nontrivial sextic-oscillator potential
(\ref{SExta}) with $q=1$ and with the binding energies
re-parametrized in accord with eq. (\ref{recipe}) where $s_1=s$,
 \ben
 E =  \frac{1}{2}\frac{g_1}{\sqrt{g_2}}\,D
 +{(64\,g_2)^{1/4}}\,\sqrt{D} \, s\,,
 \een
full attention must be paid to the selfconsistency problem
represented by the set of equations (\ref{trap}). At every $N$, its
first nontrivial $q=1$ version
 \begin{equation}
 \label{trapegy}
 \left( \begin{array}{cccccc}
 s& 1& &&&\\
 N-1 & s &2 & && \\
  & N-2 & s & 3 &&\\
 &  &\ddots & \ddots & \ddots & \\
& &&2& s & N-1     \\&& && 1 & s
\end{array} \right)
 \left( \begin{array}{c}
 {p}_0\\
 {p}_1\\
\vdots \\
 {p}_{N-2}\\
 {p}_{N-1}
\end{array} \right)
=
0\
 \end{equation}
has the form of an asymmetric eigenvalue problem. In standard manner
it leads to the secular equation (\ref{secular}) expressible as the
following sequence of the polynomial conditions,
 \ben
s^3-4\,s=0, \ \ \ \ \ \ \ N = 3,
 \een
 \ben
s^4-10\,s^2+9=0, \ \ \ \ \ \ \ N = 4,
 \een
 \ben
s^5-20\,s^3+64\,s=0, \ \ \ \ \ \ \ N = 5,
 \een
etc. By mathematical induction, all the infinite hierarchy of these
equations has been recently derived and solved in ref. \cite{RS}.

Quite remarkably, all of the {\em real} (i.e., ``physical") energy
roots $ s=s^{(j)}$ proved to be equal to integers. Moreover, all of
them may be determined by the single and compact formula
 \be
 s=s^{(j)}=-N-1+2j,\ \ \ \ j = 1, 2, \ldots, N.
 \label{sextar}
 \ee
This represented one of the key motivations of our present work,
especially when we imagined that also all the related coefficients
$p_n^{(j)}$ may equally easily be normalized to integers,
 \ben
 p_0^{(1)}=1, \ \ \ \ \ \ \ \ \ N = 1,
 \een
 \ben
 p_0^{(1)}=
 p_1^{(1)}=
 p_0^{(2)}=
 - p_1^{(2)}=1,
 \ \ \ \ \ \ \ \ \ N = 2,
 \een
 \ben
 p_0^{(1)}=
 p_2^{(1)}=
 p_0^{(2)}=
 - p_2^{(2)}=
 p_0^{(3)}=
 p_2^{(3)}=1,\ \ \
 p_1^{(1)}=
 - p_1^{(3)}=2,
 \ \ \
 p_1^{(2)}=0, \ \ \  \ N = 3,
 \een
etc.

The first result of our subsequent computations using the symbolic
manipulation techniques proved equally encouraging since we succeeded
in compactification of the set of the above recurrent solutions to
the single leading-order form of the related wave functions,
 \ben
 \psi^{(j)}(r) =r^{\ell+1}\,
 \left(
 1+\frac{r^2}{\mu}
 \right )^{N-j}\,
 \left(
 1-\frac{r^2}{\mu}
 \right )^{j-1}\,
 \exp \left (- \frac{1}{2}{\alpha}_0 r^2 -
  \frac{1}{4}{\alpha}_1 r^4
 \right )\,, \een
 \be\ \ \ \ \ \ \ \ \ \ \ \
 \ \ \ \ \ \ \ \ \
 \ \
 \ \ \ \ \
 \ \ \ \ \ j= 1, 2, \ldots, N\,.
  \label{anesex}
 \ee
A few more comments may be added.

\begin{itemize}

\item
The large and  degenerate nodal zeros in eq. (\ref{anesex}) are a
mere artifact of the zero-order construction.

\begin{itemize}

\item
The apparently interesting exact summability of all the separate
${\cal O}(r^2/\mu)$ error terms is not too relevant, indeed. Although
it leads to the zero-order nodes at $r = {\cal O}(\sqrt{\mu})= {\cal
O}(D^{1/4})$, these nodes have no real physical meaning.

\end{itemize}

\item
The leading-order perturbative approximation provides a reliable
information about the energies.

\begin{itemize}

\item
They are asymptotically degenerate, due to the large overall shift of
the energy scale as explained in section \ref{reasons}.

\item
The next-order corrections may be easily obtained by the recipes of
the textbook perturbation theory.

\item
As long as the coefficients $p_n$ are defined in integer arithmetics,
the latter strategy gives, by construction, all the above-mentioned
energy corrections without any rounding errors in a way outlined in
more detail in ref. \cite{RS}.

\end{itemize}

\end{itemize}

 \noindent
In the other words, we may say that formula (\ref{anesex}) may either
be truncated to its leading-order form $ \psi^{(j)}(r) =r^{\ell+1}\,
\exp \left (- \lambda^{(2)}(r) \right )$ or, better, its full form
may be used as a generating function which facilitates the explicit
evaluation of the coefficients $p_n^{(j)}$. In comparison with the
oversimplified harmonic oscillator, the $q=1$ wave functions may be
characterized by the similar coordinate dependence which becomes
spurious (i.e., dependent on the selected normalization) everywhere
beyond the perturbatively accessible domain of $r$.

At the same time, the energies specified by eq. (\ref{sextar}) form
and absolutely amazing multiplet. On the background of its existence,
a natural question arises whether some similar regularities could
also emerge at some of the larger integer indices $q> 1$. We are now
going to demonstrate that in spite of the growth of the technical
obstacles in dealing with the corresponding key equation
(\ref{trap}), the answer is, definitely, affirmative.

\subsection{The first generalization: Decadic oscillators with $q=2$
and any $N$}

The decadic anharmonic oscillator exhibits certain solvability
features which motivated its deeper study in non-Hermitian context
\cite{decadic}. The changes of variables make this oscillator very
closely related to the common quartic problem with a recognized
relevance of both its non-Hermitian \cite{BB} and Hermitian
\cite{Ushveridze,Dubna,quartic} QES constructions.

Paying attention to the $D \gg 1$ domain and abbreviating the
parameters $s_1=s$ and $s_2=t$ of the respective decadic-oscillator
energy and coupling in eq. (\ref{recipe}), we arrive at the
four-diagonal version of our solvability condition (\ref{trap}) at
$q=2$,
 \begin{equation}
 \label{trapegyse}
 \left( \begin{array}{ccccccc}
 s& 1& &&&&\\
 t & s &2 & &&& \\
   N-1 & t& s & 3 &&&\\
 &  N-2 & t& s & 4 &&\\
 & & \ddots &\ddots & \ddots & \ddots & \\
& &&3&t& s & N-1
 \\&&& & 2 & t&s
 \\&&& && 1 & t
\end{array} \right)
 \left( \begin{array}{c}
 {p}_0\\
 {p}_1\\
\vdots \\
 {p}_{N-2}\\
 {p}_{N-1}
\end{array} \right)
=
0\ .
 \end{equation}
This is the first really nontrivial equation of the class
(\ref{trap}). In order to understand its algebraic structure in more
detail, let us first choose the trivial case with $N=2$ and imagine
that the resulting problem
 \ben
 \left( \begin{array}{cc}
 s& 1\\
 t & s \\
   1 & t
\end{array} \right)
 \left( \begin{array}{c}
 {p}_0\\
 {p}_1
\end{array} \right)
=
0\
 \een
(with $p_1 \neq 0$ due to the definition of $N$) may  be solved by
the determination of the unknown ratio of the wavefunction
coefficients $p_0/p_1=-t$ from the last line, and by the subsequent
elimination of $t= 1/s$ using the first line. The insertion of these
two quantities transforms the remaining middle line into the cubic
algebraic equation $ s^3=1$ with the single real root $s=1$.

The next equation at $N=3$ is still worth mentioning because it shows
that the strategy accepted in the previous step is not optimal.
Indeed, in
 \be
 \left( \begin{array}{ccc}
 s& 1&0\\
 t & s &2\\
 2&t & s\\
 0&  1 & t
\end{array} \right)
 \left( \begin{array}{c}
 {p}_0\\
 {p}_1\\
 {p}_2
\end{array} \right)
=
0\ \label{ugly}
 \ee
the same elimination of $p_1/p_2=-t$ and of $p_0/p_2= (t^2-s)/2$ from
the third line leads to the apparently ugly result
 \ben
 \ba
 st^2-s^2-2t=0,\\
 t^3-3st+4=0 .\ea
 \een
Still, one should not feel discouraged, at least for the following
two reasons. Firstly, an alternative strategy starting from the
elimination of $p_0$ and $p_2$ leads to the much more symmetric pair
of the conditions
 \ben
 \ba
 t^2-s^2t+2s=0,\\
 s^2-t^2s + 2t=0 \ea
 \een
the respective pre-multiplication of which by $t$ and $s$ gives the
difference $t^3=s^3$. This means that $t=\varepsilon\,s$ where the
three eligible proportionality constants exist such that
$\varepsilon^3=1$. Thus, our problem degenerates to a quadratic
equation with the pair of the real roots $s = t = s^{(1,2)}$ such
that
 \ben
 s^{(1)}=2, \ \ \ \ s^{(2)}=-1.
 \een
The second reason for optimism is even stronger: The ``ugliness" of
the results of the elimination may be re-interpreted as an
inessential intermediate stage of the solution of eq. (\ref{ugly})
for all its four unknowns (we may always put $p_N=1$) by the
``brute-force" symbolic manipulations on the computer, by the so
called technique of Gr\"{o}bner bases \cite{onitri}. In particular,
at $N=3$, the computerized experiment of this type leads to the
step-by-step elimination of the redundant unknowns and to the final
effective polynomial equation for the single unknown quantity $s$,
 \ben
 s^6-7\,s^3-8=0.
 \een
This equation may be verified to possess the same complete set of the
real roots as above. One may conclude that the real energies of the
``strongly spiked" decadic oscillator are very easily determined even
without a detailed specification of an ``optimal" elimination
pattern.

We see that in general one may expect that eq. (\ref{trapegyse}) may
give many unphysical complex roots as well. This is confirmed by the
next step with $N=4$ leading to the effective polynomial equation
 \ben
 s^{10} - 27\,s^7 + 27\,s^4 - 729\,s = 0\,.
 \een
Being tractable by the standard computer software, it results in the
set of the mere two real roots again,
 \ben
 s^{(1)}=3, \ \ \ \ s^{(2)}=0,
   \ \ \ \ \ \ \ N = 4, \ \ \ q = 2.
  \een
It is not difficult to continue along the same path. In general one
finds that the $q=2$ problem may be reduced to a single polynomial
equation with $\left ( \ba N+1\\2 \ea \right )$ complex roots $s$.
Still, originally, we were unable to suspect that after all the
explicit calculations, {\em all} the general physical (i.e., real)
spectrum of energies proves to be quite rich and appears described
again by the following {\em closed and still almost trivial} formula
 \be
 s^{(j)}=N+2-3j, \ \ \ \ \ \ j = 1, 2, \ldots,
 j_{max}, \ \ \ \ \ \ \ j_{max}=
 entier \left [ \frac{N+1}{2}
 \right ]\ .
 \label{empir}
 \ee
This is our first important conclusion. After one applies the same
procedure at the higher and higher dimensions $N$, some more advanced
symbolic manipulation tricks must be used \cite{onitri}.
Nevertheless, one repeatedly arrives at the confirmation of the
$N-$independent empirical observation (\ref{empir}) and extends it by
another rule that at all the values of the dimension $N$, there exist
only such real roots that $s^{(j)}=t^{(j)}$. This means that each
''solvability admitting" real energy $s$ requires, purely
constructively, the choice of its own ''solvability admitting" real
coupling constant $t$.

This is our second important result which parallels completely the
similar observations made in our preceding paper on the quartic
oscillators \cite{quartic}. Now, a fully open question arises in
connection with all the $q>2$ versions of eq. (\ref{trap}). Do their
real roots $s, t, \ldots$ exhibit the similar pattern as emerged at
$q=2$?

\subsection{The second generalization: Oscillators with $q=3$
and their solvability at any $N$}

At $q=3$ we have to solve the five-diagonal eq. (\ref{trap}),
 \begin{equation}
 \label{trapegydr}
 \left( \begin{array}{cccccccc}
 r& 1& &&&&&\\
 s& r& 2& &&&&\\
 t & s &r&3 & &&& \\
   N-1 & t& s & r&4 &&&\\
 &  N-2 & t& s &r&5 &&\\
 & & \ddots &\ddots &\ddots & \ddots & \ddots & \\
 &&&4&t& s &r& N-1
 \\
 &  &&&3&t& s & r
 \\&&&& & 2 & t&s
 \\&&&& && 1 & t
\end{array} \right)
 \left( \begin{array}{c}
 {p}_0\\
 {p}_1\\
\vdots \\
 {p}_{N-2}\\
 {p}_{N-1}
\end{array} \right)
=
0\
 \end{equation}
which may be reduced, by means of the similar symbolic computations
as above, to the single polynomial problem
 \ben
 t^{9}-12\,t^{5}-64\,t=0
 \een
at $N=3$, to the next similar condition
 \ben
 t^{16}-68\,t^{12}-442\,t^8-50116\,t^4+50625=0
 \een
at $N=4$, to the conditions of vanishing of the secular polynomial
 \ben
 t^{25}-260\,t^{21}+7280\,t^{17}-1039040\,t^{13}-152089600\,t^9
 +2030239744\,t^5+10485760000\,t
 \een
at $N=5$, or to the perceivably longer equation
 \ben
t^{36}-777\,t^{32}+135716\,t^{28}-17189460\,t^{24}-3513570690\,t^{20}
 -\een \ben-1198527160446\,t^{16}
+103857100871252\,t^{12}+873415814269404\,t^8+\een
\ben+74500845455535625\,t^4 -75476916312890625=0
 \een
at $N=6$ etc. These computations represent a difficult technical task
but at the end they reveal again a clear pattern in the structure of
the secular polynomials as well as in their solutions. One arrives at
the similar final closed formulae as above. Now one only deals with
more variables so that we need two indices to prescribe the complete
classification scheme
 \ben
 s=s^{(j)}= N+3-4j, \ \ \ \ \ \ \ \
 \een
 \be
 r =
 r^{(j,k)}=t =
 t^{(j,k)}= -N-3+2j+2k,
 \ee
 \ben
  \ \ \ \ \ \ k = 1, 2, \ldots,
 k_{max}(j), \ \ \ \ \ \ k_{max}(j)=
 N+2-2j\ ,
 \een
 \ben
 \ \ \ \ \ \ \ \ \ \
 j = 1, 2, \ldots,
 j_{max}\ ,
  \ \ \
  \ \ \ \ j_{max}=
 entier \left [ \frac{N+1}{2}
 \right ]\ .
 \een
We may re-emphasize that all the real roots share the symmetry $r=t$
but admit now a different second root $s$. The physical meaning of
these roots is obvious. Thus, the energies of the oscillations in the
polynomial well
 \ben
 V^{(q=3,k=1)}(r) =  a\,r^2+b\,r^4+\ldots + g\,r^{14}
 \een
will be proportional to the roots $ r^{(j,k)}$. After the change of
variables, the roots $s^{(j)}$ will represent energies for the
alternative, ``charged" polynomial potentials
 \ben
 V^{(q=3,k=2)}(r) = \frac{e}{r}+ a\,r+b\,r^2+\ldots + f\,r^6
 \een
etc \cite{classif}.

\section{Outlook: QES solutions at $q\geq 4$ and selected $N$}

\subsection{An apparent loss of simplicity at $q=4$ and $N \leq 6$}

In our present formulation of the problem (\ref{trap}), we denote the
descending diagonals as $s_{m}$ with $m=1,2,3,4$ and get the equation
 \begin{equation}
 \label{tregydr}
 \left( \begin{array}{cccc}
 s_1& 1& &\\
 s_2& \ddots&\ddots& \\
  s_3&\ddots&\ddots&N-1\\
 s_4&\ddots&\ddots& s_1\\
 N-1&\ddots&\ddots&s_2\\
 &\ddots&\ddots&s_3\\
 &&1&s_4
\end{array} \right)
 \left( \begin{array}{c}
 {p}_0\\
 {p}_1\\
\vdots \\
 {p}_{N-1}
\end{array} \right)
=
0\ .
 \end{equation}
Its systematic solution does not parallel completely the
above-described procedures. In fact, the reduction of the problem to
the search for the roots of a single polynomial secular equation
$P(x)=0$ (in the selected auxiliary variable $x = -s_4$) enables us
only to factorize $P(x)$ on an extension of the domain of integers,
  \ben
  P(x)= \left (x+3\right )
 \left (2\,x+1-\sqrt
{5}\right ) \left (2\,x+1+\sqrt {5}\right )
 \een
 \ben
  \left (2\,{x}^{2}-3\,x+3\,\sqrt
{5}x+18\right ) \left (2\,{x}^{2}-3\,x-3\,\sqrt {5}x+18\right )
 \een
 \ben
 \left
(2\,{x}^{2}-3\,x- \sqrt {5}x+8+2\,\sqrt {5}\right )\left
(2\,{x}^{2}-3\,x+\sqrt {5}x+8-2 \,\sqrt {5}\right )
 \een
 \ben
 \left
({x}^{2}+x+\sqrt {5}x+4+\sqrt {5}\right )
  \left ({x}^{2}+x-\sqrt
{5}x+4-\sqrt {5}\right )
 \een
 \ben
 \left (-2\,\sqrt {5}+8- 3\,x+3\,\sqrt
{5}x+2\,{x}^{2}\right )
 \left (2\,\sqrt {5}+8-3\,x-3\, \sqrt
{5}x+2\,{x}^{2}\right )
 \een
 \ben
 \left (-2\,\sqrt {5}+8+7\,x-\sqrt {5}x+2
\,{x}^{2}\right )
 \left (2\,\sqrt {5}+8+7\,x+\sqrt {5}x+2\,{x}^{2}
\right )
 \een
 \ben
 \left (2\,{x}^{2}+2\,x+3-\sqrt {5}\right )
 \left (2\,{x}^{2}+2
\,x+3+\sqrt {5}\right )
 \een
 \ben
 \left (\sqrt {5}+3-3\,x-\sqrt {5}x+2\,{x}^{2}
\right )\left (-\sqrt {5}+3-3\,x+\sqrt {5}x+2\,{x}^{2}\right )
 \een
 \ben
 \left
(2 \,\sqrt {5}+8-3\,x+\sqrt {5}x+2\,{x}^{2}\right )
 \left (-2\,\sqrt
{5}+8 -3\,x-\sqrt {5}x+2\,{x}^{2}\right )\ .
 \een
From this lengthy formula it follows that we get
 \ben
 s_4^{(1)}=3, \ \ \ \
 s_4^{(2)}=\frac{\sqrt{5}+1}{2} \approx 1.618, \ \ \ \
 s_4^{(3)}=\frac{\sqrt{5}-1}{2} \approx -0.618\ .
 \een
There only exist these three real roots $s_4$ in this case.

The similar computerized procedure gave us the real roots also at
$N=5$. Their inspection leads to the conclusion that $s_2=s_3$,
$s_1=s_4$. We did not succeed in an application of our algorithms
beyond $N=5$ yet. Even the $N=5$ version of eq. (\ref{tregydr}) in
its reduction to the condition
 \ben
 x^{70}-936\,x^{65}+67116\,x^{60}-95924361\,x^{55}-74979131949\,x^{50}
 +8568894879002\,x^{45}-
 \een
 \ben
 \ldots -17459472274501870222336\,x^5+142630535951654322176=0
 \een
of the vanishing auxiliary polynomial required a fairly long
computation for its (still closed and compact) symbolic-manipulation
factorization summarized in Table~1.

Marginally, it is worth noticing that the choice of $q=4$ is the
first instance where the popular cubic polynomial forces may emerge
via the change of variables of ref. \cite{classif}. For this reason,
in particular, the incomplete character of our $q=4$ solution might
prove challenging in the context of the so called PT symmetric
quantum mechanics where the study of cubic oscillators happened to
play a particularly significant role \cite{ix3}.

\subsection{Simplicity re-gained at $q=5$}

\subsubsection{$N=6$}

At $q=5$ and $N=6$ the symbolic manipulations using the Gr\"{o}bner
bases \cite{Groebner} generate the secular polynomial in $x=s_5$
which has the slightly deterring form
 \ben
x^{91}-16120\,x^{85}+49490694\,x^{79}-286066906320\,x^{73}-3553475147614293\,x^{67}-
 \een
 \ben
 \ldots
 -319213100611990814833843025405983064064000000\,x=0\ .
  \een
Fortunately, it proves proportional to the polynomial with the mere
equidistant and simple real zeros,
 \ben
 P_1^{(6)}(x)=
x\left (x^2-1\right )\left (x^2-2^2\right )\left (x^2-3^2\right
)\left (x^2-4^2 \right )\left (x^2-5^2\right )\ .
 \een
The rest of the secular polynomial is a product of the other two
elementary and positive definite polynomial factors
 \ben
 P_2^{(6)}(x)=\prod_{k=1}^{2}\,
 \left (
 x^2-3k\,x+3k^2
 \right )
 \left (
 x^2+3k^2
 \right )
 \left (
 x^2+3k\,x+3k^2
 \right )
 \een
and
 \ben
 P_{3}^{(6)}=
 \prod_{k=1}^{5}\,
 \left (
 x^2- k\,x+k^2
 \right )
 \left (
 x^2+ k\,x+k^2
 \right ),
 \een
with another positive definite polynomial
 \ben
 P_{4}^{(6)}=
 \prod_{k=1}^{12}\,
 \left (
 x^2- b_k\,x+c_k
 \right )
 \left (
 x^2+ b_k\,x+c_k
 \right )\
 \een
where the structure of the two series of coefficients (see their list
in Table~2) is entirely enigmatic.

The subsequent symbolic manipulations reveal a symmetry $s_2=s_4$ and
$s_1=s_5$ of all the real eigenvalues. In the pattern summarized in
Table~3, we recognize a clear indication of a return to the
transparency of the $q \leq 3$ results which may be written and
manipulated in integer arithmetics.

\subsubsection{$N=7$}

One should note that in spite of its utterly transparent form, the
latter result required a fairly long computing time for its
derivation. One encounters new technical challenges here, which will
require a more appropriate treatment in the future \cite{mytri}.
Indeed, the comparison of the $N=6$ secular polynomial equation with
its immediate $N=7$ descendant
 \ben
x^{127}-60071\,x^{121}+1021190617\,x^{115}-11387407144495\,x^{109}-\ldots+
c\,x\cdot
 10^6 =0
 \een
shows that the last coefficient
 \ben
c= 125371220122726667620073789326658415654595883041274311330630729728
 \een
fills now almost the whole line. This case failed to be tractable by
our current computer code and offers the best illustration of the
quick growth of the complexity of the $q\geq 5$ constructions with
the growth of the QES dimension parameter $N$.

Fortunately, we are still able to keep the trace of the pattern
outlined in Tables~2 and~3. Indeed, our new secular polynomial
factorizes again in the product of four factors $P_j(x)$, $j = 1, 2,
3, 4$ where only the first one has the real zeros,
 \ben
 P_1^{(7)}(x)=P_1^{(6)}(x)\cdot
\left (x^2-6^2\right )\ .
 \een
The further three factors fit the structure of their respective
predecessors very well,
 \ben
 P_2^{(7)}(x)=P_2^{(6)}(x)\cdot
 \left (
 x^2-9\,x+27
 \right )
 \left (
 x^2+27
 \right )
 \left (
 x^2+9\,x+27
 \right )
 \een
and
 \ben
 P_{3}^{(7)}=P_{3}^{(6)}\cdot
  \left (
 x^2- 6\,x+36
 \right )
 \left (
 x^2+ 6\,x+36
 \right )
 \een
while
 \ben
 P_{4}^{(7)}=P_{4}^{(6)}\cdot
 \prod_{k=1}^{6}\,
 \left (
 x^2- f_k\,x+g_k
 \right )
 \left (
 x^2+ f_k\,x+g_k
 \right )\ .
 \een
The subscript-dependence of the new coefficients is listed in
Table~4.

On the basis of the above factorization we may deduce that the
pattern of Table~2 survives, {\it mutatis mutandis}, also the
transition to $N=7$. Indeed, by inspection of Tables~2 and~4 one
easily proves that the product function $P_2(x)\,P_3(x)\,P_4(x)$ has
no real zeros and remains positive on the whole real line of $x$
again. A full parallel with the $N=6$ pattern is achieved and might
be conjectured, on this background, for all $N$, therefore.


\section{Summary}

Our paper offered new closed solutions of Schr\"{o}dinger equation
with polynomial potentials at the large angular momenta $\ell \gg 1$.
This type of construction proves well founded and motivated, say, in
nuclear physics where, quite naturally, the variational calculations
in hyperspherical basis lead to the very large values of $\ell =
{\cal O}(10^3)$ \cite{Sotona}. In such a setting, of course,
practically {\em any} version of the popular $1/\ell$ perturbation
expansion (a compact review may be found, e.g., in papers
\cite{Bjerrum}) must necessarily lead to a satisfactory numerical
performance.

Our present project was more ambitious. We imagined that a rarely
mentioned {week point} of all the above perturbative philosophy lies
in the notoriously narrow menu of the necessary zero-order
approximants $H_0$ \cite{Marcelo}. Indeed, in spite of an amazing
universality of all the different $1/\ell$ (better known as $1/N$)
expansion techniques (cf. a small sample of the relevant
{computational} tricks in \cite{jednaden}), one usually finds and
returns to the common harmonic oscillator $H_0^{(HO)}\equiv
H^{(q=0)}$, in spite of the wealth and variability of the underlying
{physics} \cite{physics}. For this reason we recently started to
study some alternative possibilities offered by the QES models
\cite{RS,quartic}. In our present continuation of this effort, a
decisive extension of the results of this type is given.

Our text reveals the existence and describes the construction of
certain {fairly large} multiplets of ``exceptional" $\ell \gg 1$
bound states for a very broad class of polynomial oscillators. We
believe that they might find an immediate application in some
phenomenological $D \gg 1$ models where the enhancement of the
flexibility of the models with $q>1$ might lead, say, to a more
precise fit of the vibrational spectra etc.

From the mathematical point of view, the most innovative and
characteristic feature of our new $D \gg 1$ QES multiplets lies in
the existence of the new {\em closed and compact} formulae for the
QES energies and/or couplings {\em at all $N$}. For this reason, the
corresponding partially solvable polynomial oscillator Hamiltonians
$H_0^{(q,N)}$ might even be understood as lying somewhere in between
the QES and ES classes.

Due to such an exceptional transparency of our constructions of
$H_0^{(q,N)}$, a facilitated return to the ``more realistic" finite
spatial dimensions $D={\cal O}(1)$ might prove tractable by
perturbation techniques. Two reasons may be given in favor of such a
strategy. First, due to the specific character of our present
``unperturbed" spectra {\em and} eigenvectors, the perturbation
algorithm might be implemented {\em in integer arithmetics} (i.e.,
without rounding errors) in a way outlined, preliminarily, in ref.
\cite{RS} at $q=1$. Second, the evaluation of the few lowest orders
might suffice. This expectation follows from the enhanced flexibility
of the available zero-order Hamiltonians. {\it A priori}, a better
convergence of the corrections will be achieved via a better
guarantee of a ''sufficient smallness" of the difference between the
realistic Hamiltonian $H$ and its available approximant $H_0$.

Of course, the detailed practical implementation of the perturbation
technique represents an independent task which must be deferred to a
separate publication. With encouraging results, the first steps in
this direction have already been performed at $q=2$ \cite{quartic}.
In parallel, it seems feasible to enlarge further the range of $q$ in
zero order. Although one has to deal with the fairly complicated
symbolic manipulations on the computer beyond $q=3$, we still intend
to perform a deeper analysis of the problems with $q\geq 4$ in the
nearest future \cite{mytri}.


\subsection*{Acknowledgements}

M. Z.  appreciates the support by the grant Nr. A 1048302 of GA AS
CR. The contribution of V. G. and D. Y. was supported in part by the
grants 00-15-96691 and 01-01-007 from Russian Foundation for Basic
Research.

\section*{Tables}


Table 1. Sample of the real roots of eq. (\ref{tregydr}) ($q=4$).
  $$
\begin{array}{||c|c||} \hline \hline
 \multicolumn{2}{||c||}{N=4}\\ \hline
 s_{3}&s_{4}\\
 \hline
3&3\\ ({\sqrt{5}+1})/{2}&({-\sqrt{5}+1})/{2}\\
({-\sqrt{5}+1})/{2}&(\sqrt{5}+1)/{2}\\ \hline \hline
 \multicolumn{2}{||c||}{N=5}\\ \hline
 s_{3}&s_{4}\\
 \hline
  -1&-1\\
4&4\\ \sqrt{5}-1&-\sqrt{5}-1\\ -\sqrt{5}-1&\sqrt{5}-1\\
({\sqrt{5}+3})/{2}&({-\sqrt{5}+3})/{2}\\
({-\sqrt{5}+3})/{2}&(\sqrt{5}+3)/{2}\\
 \hline \hline
 \multicolumn{2}{||c||}{N=6}\\ \hline
 s_{3}&s_{4}\\
 \hline 0&0\\ 5&5\\ \sqrt{5}
&-\sqrt{5}
\\ -\sqrt{5}&\sqrt{5} \\ ({\sqrt{5}
+5})/{2}&({-\sqrt{5} +5})/{2}\\ ({-\sqrt{5} +5})/{2}&({\sqrt{5}
+5})/{2}\\ \hline \hline
\end{array}
$$


Table 2. Coefficients $b_k$ and $c_k$ for $q=5$ and $N=6$.
 $$
 \begin{array}{||r|r|r||}
\hline  \hline  k & c_k & b_k
\\ \hline
1 - 3& 7& 1, 4, 5\\
 4 - 6& 13& 2, 5, 7\\
 7 - 9& 19& 1, 7, 8\\
 10 - 12& 21 & 3, 6, 9
\\ \hline \hline
\end{array}
$$


Table 3. Real roots of eq. (\ref{trap}) at $q=5$ and $N=6$
 $$
 \begin{array}{||rrr||}
\hline  \hline  s_3 & s_4 & s_5
\\ \hline -5&5&-5\\ -3&5&-3\\ -1&5&-1\\ 1&5&1\\ 3&5&3\\ 5&5&5\\
\hline -5&-1&1\\ -3&-1&3\\ -1&-1&-1\\ 1&-1&1\\ 3&-1&-3\\ 5&-1&-1\\
\hline -5&2&-2\\ -3&2&0\\ -1&2&-4\\ -1&2&2\\ 1&2&-2\\ 1&2&4\\ 3&2&0\\
5&2&2\\ \hline \hline
\end{array}
$$


Table 4. Additional coefficients $f_k$ and $g_k$ at $q=5$ and $N=7$.
 $$
 \begin{array}{||r|r|r||}
\hline \hline  k & f_k & g_k
\\ \hline
 1 - 3& 28& 2, 8, 10\\
 4 - 6& 31& 4, 7, 11
\\ \hline \hline
\end{array}
$$


\begin{thebibliography}{00}

\bibitem{Hautot}
Hautot A 1972 Phys. Lett. A 38  305

\bibitem{Singh}
Singh V, Biswas S N and  Datta K 1978 Phys. Rev. D 18 1901

\bibitem{Flessas}
Flessas G P 1979 Phys. Lett. A 72 289

\bibitem{Magyari}
Magyari E 1981 Phys. Lett. A 81 116

\bibitem{singular}
Znojil M 1982 J. Phys. A: Math. Gen. 15 2111

\bibitem{Turbiner}
Turbiner A 1988 Commun. Math. Phys. 118 467;

Shifman M A 1989 Int. J. Mod. PHys. A 4 2897

\bibitem{Miller}
Miller W Jr. 1968
              Lie theory of special functions,
          (New York: Academic)

\bibitem{Ushveridze}
Ushveridze A G 1994 Quasi-Exactly Solvable Models in Quantum
Mechanics (Bristol: IOPP)

\bibitem{Winternitz}
Fri\v{s} J, Mandrosov V, Smorodinsky Ya A,  Uhl\'{\i}\v{r} M and
Winternitz P 1965 Phys. Lett. 16 354;

Tempesta P, Turbiner A V and Winternitz P 2001 J. Math. Phys. 42 4248

\bibitem{Fluegge3}
Fl\"{u}gge S 1971 Practical quantum mechanics I (Berlin: Springer),
p. 168;

Grosche C, Pogosyan G S, Sissakian A N 1995 Fortsch. Phys. 43 453

\bibitem{BG}
Buslaev V and Grecchi V 1993 J. Phys. A: Math. Gen. 26 5541

\bibitem{FluLag}
Abramowitz M and Stegun I A 1970
               Handbook of Mathematical Functions
               (New York: Dover)

\bibitem{CKS}
Olver F W J 1974 Introduction to Asymptotics and Special Functions
(New York: Academic), ch. 6;

Cooper F, Khare A and Sukhatme U 1995 Phys. Rep. 251 267

\bibitem{Leach}
Znojil M and Leach P G L 1992 J. Math. Phys. 33 2785

\bibitem{classif}
Znojil M 1994 J. Phys. A: Math. Gen. 27 4945

\bibitem{Dubna}
Znojil M 1989 Anharmonic oscillator in the new perturbative picture
(Dubna: JINR) report Nr. E5 - 89 - 726

\bibitem{RS}
Znojil M 2002 Proc. Inst. Math. NAS (Ukraine) 43 777

\bibitem{decadic}
Znojil M 2000 J. Phys. A: Math. Gen. 33  6825

\bibitem{BB}
Bender C M and Boettcher S 1998 J. Phys. A: Math. Gen. 31 L273;

Bender C M and Milton K A 1998 Phys.Rev. D 57 3595;

Znojil M 1999 J. Phys. A: Math. Gen. 32  7419;

Znojil M 2000 J. Phys. A: Math. Gen. 33  4203;

Znojil M, Cannata F, Bagchi B and Roychoudhury R 2000 Phys. Lett. B
483 284

\bibitem{quartic}
Znojil M 1999 J. Math. Chem. 26  157

\bibitem{onitri}
Sturmfels B 2002 Solving Systems of Polynomial Equations (Providence:
AMS)

\bibitem{ix3}
Caliceti E, Graffi S and Maioli M 1980 Commun. Math. Phys. 75 51;

Bender C M and Boettcher S 1998 Phys. Rev. Lett. { 24}  5243;

Fern\'andez F M, Guardiola R,  Ros J and Znojil M 1998 J. Phys. { A}:
Math. Gen. 31 10105;

Dorey P, Dunning C and Tateo R 2001 J. Phys. A: Math. Gen. 34 5679

\bibitem{Groebner}
Adams W W and Loustaunau P 1991 An Introduction to Gr\"{o}bner Bases
(Providence: AMS)

\bibitem{mytri}
Yanovich D, Gerdt V and Znojil M, in preparation

\bibitem{Sotona}
Sotona M and \v{Z}ofka J 1974 Phys. Rev. C 10 2646;

Lombard R J and Mare\v{s} J 1999 Phys. Rev. D 59 076005

\bibitem{Bjerrum}
Bjerrum-Bohr N E J 2000 J. Math. Phys. 41 2515

\bibitem{Marcelo}
Fern\'{a}ndez F M 2001 Introduction to Perturbation Theory in Quantum
Mechanics (Boca Ranton: CRC Press)

\bibitem{jednaden}
Imbo T, Pagnamenta A and Sukhatme U 1984 Phys. Rev. D 29 1669;

Varshni Y P 1987 Phys. Rev. A 36 3009;

Fernandez F M 2002 J. Phys. A: Math. Gen. 35 10663;

Mustafa O  2002 J. Phys. A: Math. Gen. 35 10671

\bibitem{physics}
Witten E 1979  Nucl. Phys. B 160 57;

Yaffe L G 1982 Rev. Mod. Phys. 54 407;

Popov V S, Sergeev A V and Shcheblykin A V 1992 Zhurnal
Experimentalnoy i Teoreticheskoy Fiziki 102 1453;

Cooper F, Habib S, Kluger Y, Mottola E,  Paz J P and Anderson P R
1994 Phys. Rev. D 50 2848;

Schiller A and K. Ingersent K 1995  Phys. Rev. Lett.  75 113;

Irkhin V Yu, Katanin A A and Katsnelson M I 1996 Phys. Rev. B 54
11953;

Bhattacharya T, Lacaze R and Morel A 1997 J. Phys. I (France) 7 1155;

Parisi G and F. Slanina F 1999 Eur. Phys. J. B 8 603

\end{thebibliography}
\end{document}